%% file: paper.tex
\documentclass[sigconf,nonacm,natbib=false]{acmart}

\RequirePackage[abbreviate=true, dateabbrev=true, isbn=true, doi=true, urldate=comp, url=false, maxbibnames=9, backref=false, backend=biber, style=ACM-Reference-Format, language=american]{biblatex}

\usepackage{booktabs} 
\usepackage{longtable} 
\usepackage{multirow}
\usepackage{listings}

\setcopyright{rightsretained}

\acmDOI{10.475/123_4}

\acmISBN{123-4567-24-567/08/06}

\acmConference[WOODSTOCK'97]{ACM Woodstock conference}{July 1997}{El
  Paso, Texas USA}
\acmYear{1997}
\copyrightyear{2016}

\acmPrice{15.00}

\begin{document}
\title{Data Provenance for Sport}

\author{Andrew J. Simmons}
\orcid{0000-0001-8402-2853}
\affiliation{%
  \institution{Deakin University}
  \department{Applied Artificial Intelligence Institute}
  \streetaddress{Locked Bag 20000}
  \city{Geelong}
  \state{VIC}
  \country{Australia}
  \postcode{3220}
}
\email{a.simmons@deakin.edu.au}

\author{Scott Barnett}
\affiliation{%
  \institution{Deakin University}
  \department{Applied Artificial Intelligence Institute}
  \streetaddress{Locked Bag 20000}
  \city{Geelong}
  \state{VIC}
  \country{Australia}
  \postcode{3220}
}
\email{scott.barnett@deakin.edu.au}

\author{Simon Vajda}
\affiliation{%
  \institution{Deakin University}
  \department{Applied Artificial Intelligence Institute}
  \streetaddress{Locked Bag 20000}
  \city{Geelong}
  \state{VIC}
  \country{Australia}
  \postcode{3220}
}
\email{s.vajda@deakin.edu.au}

\author{Rajesh Vasa}
\affiliation{%
  \institution{Deakin University}
  \department{Applied Artificial Intelligence Institute}
  \streetaddress{Locked Bag 20000}
  \city{Geelong}
  \state{VIC}
  \country{Australia}
  \postcode{3220}
}
\email{rajesh.vasa@deakin.edu.au}


\begin{abstract}
Data analysts often discover irregularities in their underlying dataset, which need to be traced back to the original source and corrected. Standards for representing data \textit{provenance} (i.e. the origins of the data), such as the W3C PROV standard, can assist with this process, however require a mapping between abstract provenance concepts and the domain of use in order to apply them effectively. We propose a custom notation for expressing provenance of information in the sport performance analysis domain, and map our notation to concepts in the W3C PROV standard where possible. We evaluate the functionality of W3C PROV (without specialisations) and the VisTrails workflow manager (without extensions), and find that as is, neither are able to fully capture sport performance analysis workflows, notably due to limitations surrounding capture of automated and manual activities respectively. Furthermore, their notations suffer from ineffective use of visual design space, and present potential usability issues as their terminology is unlikely to match that of sport practitioners. Our findings suggest that one-size-fits-all provenance and workflow systems are a poor fit in practice, and that their notation and functionality need to be optimised for the domain of use.
\end{abstract}

%
%



\maketitle

\input{body}

\printbibliography

\input{appendix}

\end{document}

%% file: body.tex
\section{Introduction}

{Sport performance analysis involves a combination of manual annotation
of video, automatable derivation of performance statistics from the
annotations, and ad-hoc interplay of manual and automated processes to
refine data and define new metrics. The competitive nature of sport, and
the the explosion of available data captured by in-game sensors, had led
to demand for increasingly sophisticated forms of analysis. However,
without some form of data }{provenance}{~describing all processes and
data sources used in the derivation of the final performance statistic,
there is limited ability to reproduce the analysis, nor to audit the
process for human error, software bugs, or data entry errors that may
have affected the result.}

{}

{We begin by providing a motivating scenario inspired by real challenges
faced by sport performance analysts, and highlight the need for data
provenance to audit and reproduce the processes. These scenarios are
used to elicit requirements, that form the basis for our proposed
provenance notation optimised for sport performance analysis.}

{}

{We then evaluate the functionality, notational effectiveness, and
usability of existing tools for the description and capture of data
provenance, specifically the W3C PROV standard and the VisTrails
workflow manager. We identify shortcomings of existing systems, and
conclude with recommendations on how to bridge the language gap between
abstract provenance concepts and the sport performance domain.}

\section{Motivating Scenario}\label{sec:provmotivation}

{Consider Ellie, a high performance sport performance analyst for an
Australian Rules Football team, who wants to test a new player
evaluation metric.}

{}

\subsection{Physical Provenance Scenario}

{Ellie begins by annotating video footage of past games using a timeline
annotation tool, such as Sportscode\footnote{\url{https://www.hudl.com/elite/sportscode}}. From the
centre bounce (start of play), Player 3 taps the ball to player 12, who
kicks it to Player 7, who scores a goal. As per the laws of the game,
after the goal, the ball is returned to the centre of the field for the
next centre bounce.}

{}

{Upon annotating the video footage from all past games, Ellie decides to
investigate one of the goals within more detail. For example, she might
want to investigate goal assists that led to scoring the goal (assume
that the club does not already have a custom label to represent the set
of goal assists). While she can re-watch the video footage, ideally she
would like to be able to extract an abstract representation of the
provenance of the goal (i.e. how the goal came to be) using the data
that she has coded in order to allow her to efficiently investigate a
large number of cases without needing to re-watch the footage.}

{}

{Within her timeline tool, Ellie is able to search for a goal and scan
back in time to see the possession chain, however her timeline is
cluttered with additional annotations such as the medical team's
annotation of an on-field injury to Player 3's knee. While she can hide
certain event types, she cannot instruct her timeline tool to
automatically hide everything that did not contribute to the goal, as
her timeline tool has no concept of how events are connected to each
other. Furthermore, she sees events prior to the centre bounce and after
the goal, as her timeline tool does not recognise that these events
reset the game state.}

{}

\subsection{Workflow Provenance Scenario}

{Ellie's timeline tool allows her to qualitatively analyse specific
events through the medium of video, but does not provide a way for her
to directly compute custom metrics from her annotations. To do so, she
exports her timeline annotations to an intermediate format (e.g. CSV),
so that she can statistically analyse the data using an external
analysis tool (e.g. Microsoft Excel).}

{}

{Prior to conducting the analysis, Ellie de-identifies the exported
annotation data }{by substituting player identifiers with}{~anonymised
codes. This allows her to collaborate on the analysis with external
researchers who for privacy reasons should not be given access to
identifiable player data. Ellie retains a private copy of the mapping
between player identifiers and anonymous codes.}

{}

{Using}{~her analysis tool, }{Ellie imports the de-identified game
annotations, and -- with some assistance from her research collaborators
-- computes the player evaluation metric for each (anonymised) player.
Once the analysis is complete, Ellie re-identifies the players in the
final output using the mapping she kept.}

{}

{Player 7 is upset at the result of their metric, and}{~requests to see
game }{video clips of events that contributed to the calculation.
Fortunately, Ellie saved the intermediate calculation spreadsheet, but
the calculations are}{~}{difficult for Ellie to explain, as the}{~the
inputs are}{~expressed as numerical time offsets rather than embedded
video clips, and furthermore the calculations were performed using
anonymised identifiers. In order to allow the player to audit the
calculations, Ellie has to reverse the process by looking up the
anonymised identifier for Player 7 such that she can find the relevant
calculations, then extract video segments for each time offset
associated with inputs to the calculations records.}

{}

{Upon scrutinising the raw video with the player, Ellie notices that the
video shows that one of the missed goals was due to high wind conditions
rather than the fault of the player, but the wind sensor (anemometer)
was malfunctioning at the time so wasn't automatically accounted for in
Ellie's model of goal opportunity. Ellie manually overrides the data in
the wind sensor file for that period to indicate high wind conditions,
and reruns her calculations. However, she has to be cautious that her
manual changes aren't overwritten when she next synchronises sensor data
with the device.}

\begin{table}
  \caption{Requirements to support tasks performed by sport performance analyst}
  \label{Tasks}
  \begin{tabular}{cc}
\toprule
\begin{minipage}[t]{0.47\columnwidth}\raggedright\strut
{\textbf{Requirement}}
\strut\end{minipage} &
\begin{minipage}[t]{0.47\columnwidth}\raggedright\strut
{\textbf{Description}}
\strut\end{minipage}\tabularnewline
\midrule
\begin{minipage}[t]{0.47\columnwidth}\raggedright\strut
{Integrated support for working with video data}
\strut\end{minipage} &
\begin{minipage}[t]{0.47\columnwidth}\raggedright\strut
{The ability to interactively annotate segments of a video timeline as
events of interest, capture the relationships between these events, and
to visually playback the video for an event. }
\strut\end{minipage}\tabularnewline
\midrule
\begin{minipage}[t]{0.47\columnwidth}\raggedright\strut
{Support for automated processes}
\strut\end{minipage} &
\begin{minipage}[t]{0.47\columnwidth}\raggedright\strut
{The ability to automate interconnected computations such that they can
be recomputed on an updated dataset with minimal manual intervention.}
\strut\end{minipage}\tabularnewline
\midrule
\begin{minipage}[t]{0.47\columnwidth}\raggedright\strut
{Support for manual interaction}
\strut\end{minipage} &
\begin{minipage}[t]{0.47\columnwidth}\raggedright\strut
{The ability to interweave manual processes with automated processes
within a workflow, and to manually override the result of automated
processes.}
\strut\end{minipage}\tabularnewline
\midrule
\begin{minipage}[t]{0.47\columnwidth}\raggedright\strut
{Partial / shared workflow graphs}
\strut\end{minipage} &
\begin{minipage}[t]{0.47\columnwidth}\raggedright\strut
{The ability to share different parts of the workflow with different
users (e.g. external collaborators should not }{be able to reverse
the}{~de-identification operation), and to merge changes from other
users (e.g. changes suggested by external collaborators) back into one's
own workflow.}
\strut\end{minipage}\tabularnewline
\midrule
\begin{minipage}[t]{0.47\columnwidth}\raggedright\strut
{Provenance / Reverse Debugging}
\strut\end{minipage} &
\begin{minipage}[t]{0.47\columnwidth}\raggedright\strut
{The ability to trace the provenance of an analysis result back to the
raw inputs that contributed, and to scrutinise the intermediate
calculations at each step of the process. }
\strut\end{minipage}\tabularnewline
\midrule
\begin{minipage}[t]{0.47\columnwidth}\raggedright\strut
{Streaming data}
\strut\end{minipage} &
\begin{minipage}[t]{0.47\columnwidth}\raggedright\strut
{The ability to perform calculations in real-time as new data become
available. To prevent latency, automated processes should be performed
in parallel where possible, and recompute only what is necessary.
Similarly, any manual processes in the workflow should be crowdsourced
to a team of annotators to prevent bottlenecks.}
\strut\end{minipage}\tabularnewline
\bottomrule
\end{tabular}
\end{table}

\subsection{Streaming Scenario}

{The coach is impressed with Ellie's proposed metric, and asks if she
could annotate the game live as it is played and provide regular updates
of each player's metric over the course of the game. While existing
timeline annotation tool interfaces provide buttons and hotkeys to allow
the data entry rate needed for annotating the game live, Ellie's current
workflow for calculating her metric requires manually exporting the data
and running a computationally intensive process. She needs a mechanism
to automatically recompute the results in real-time as new data become
available.}

{}

\subsection{Requirements Elicitation}

{From the pain points outlined in the above tasks, we extract
requirements for the solution. These are presented in Table
\ref{Tasks}.}

\section{Background}

{Sport performance analysis is a form of applied sport science, and
implicitly involves the construction of scientific workflows to analyse
data (note that workflows can involve ad hoc human tasks, and are not
necessarily formally documented, if at all).}

{}

{Scientific workflows \cite{Gil2007} may involve both manual and
automated processes, as well as ad hoc data transformations to explore
the data from different perspectives \cite{Jankun-Kelly2007}. It is
generally accepted that one should, in principle, be able to reproduce
the steps in order to obtain the same final result. In practice however,
science is facing a ``reproducibility crisis'' \cite{Baker2016} wherein
researchers are unable to reproduce others' results, or in many cases
their own. }{Data provenance }{systems aim to alleviate this issue
through support for capture and query of information pertaining to the
origins of data, such as the primary data source, processes
applied, and agents (i.e. both humans and software) involved.}

{}

{While systems for automated workflows and provenance capture have
gained traction in specialised domains such as bioinformatics, the use,
or indeed recognition of the need for provenance more generally, such as
in the biomedical field as a whole remains ``quite low''
\cite{Baum2017}.}

{}

{Prominent scientific workflow management tools include VisTrails
\cite{Callahan2006}, Taverna \cite{Oinn2004}, and Kepler \cite{Bowers2006}.
VisTralils and Taverna represent the workflow of tasks as a directected
acyclic graph (DAG), while Kepler provides the user with a choice of the
model of computation that will be used. Workflow systems can be
integrated with data provenance systems in order to capture both the
process (prospective provenance) and trace of results (retrospective
provenance) \cite{Missier2013}.}

{}

{The W3C PROV standard \cite{Moreau2015} was introduced in 2013 in an
attempt to standardise }{provenance sharing on the Web. The PROV
standard is a component of the semantic web that cross-cuts the
ontology, logic and proof layer of the semantic web \cite{Moreau2013} (note that these layers
were part of the semantic web vision, but some aspects, particularly the
poof layer, remain ``largely unrealized'' \cite{Shadbolt2006}). Since its
release, PROV has been proposed for a range of applications including
tracking the source of citation information in curated citation
databases \cite{Peroni2017}, as an export format for Git version control
history \cite{DeNies2013}, and as a tool for coordination of human and
autonomous agents in disaster response\footnote{\url{http://www.orchid.ac.uk/}}. VisTrails
and Taverna both support export of data provenance information according
to the W3C PROV standard.\footnote{\url{https://github.com/taverna/taverna-prov}}
\footnote{``PROV support'' \url{https://github.com/VisTrails/VisTrails/issues/1075}}}

{}

{According to the W3C PROV specification, entities may be ``physical,
digital, conceptual \ldots{} real or imaginary''
~\footnote{\url{https://www.w3.org/TR/2013/REC-prov-o-20130430/\#Entity}}. This has led others to consider the use of
the specification as a means to model physical provenance, such as the
process of creating scientific specimens \cite{Cox2015}, and as a tool
for modelling the provenance of food \cite{Batlajery2018} to infer
sources of contamination. When modelling the provenance of physical
systems in this manner, provenance is often assigned a causal definition
(i.e. arrows represent causality rather than just dependency), which may
optionally be supplemented with probabilities to permit Bayesian
reasoning using the provenance graph \cite{Chapman2010}.}

\section{Approach}

{}

\subsection{Physical Provenance}

{}

{In this section, we consider the suitability of the W3C PROV
specification as a tool to model in-game sports events. Specifically, we
focus on modelling the physical provenance of the ball (i.e. the game
states that it transitions through). We achieve this through the
following mapping of concepts in the sport domain to concepts in the W3C
provenance standard: game states (i.e. position on the field and state
of possession) as PROV \textit{entities}; actions that transform the game
state (e.g. kicks) as PROV \textit{activities}; and players that perform the
actions as PROV \textit{agents}. To support reasoning about the game in terms
of either specific players (e.g. Cyril Rioli) or the roles they
represent (e.g. Half Forward), we use the PROV
\textit{actedOnBehalfOf} relation to describe a many:many relationship
between players and roles. This allows our model to handle role changes
(e.g. a substitution of player roles due to an injury).}

{}

{While this mapping is sufficient for formalisation purposes, we must
also consider the usability of such a system by a sport performance
analyst. Specifically, the abstract concepts of entities, activities and
agents are unlikely to be familiar to users in the sport domain, and
thus breaks the usability heuristic that software should ``speak the
user's language'' \cite{Nielsen1994}. As such, we propose specialising
the notation of PROV with custom symbols for game events in order to
translate it into the language of sport.

\begin{figure}[!htb]
\includegraphics[width=\columnwidth]{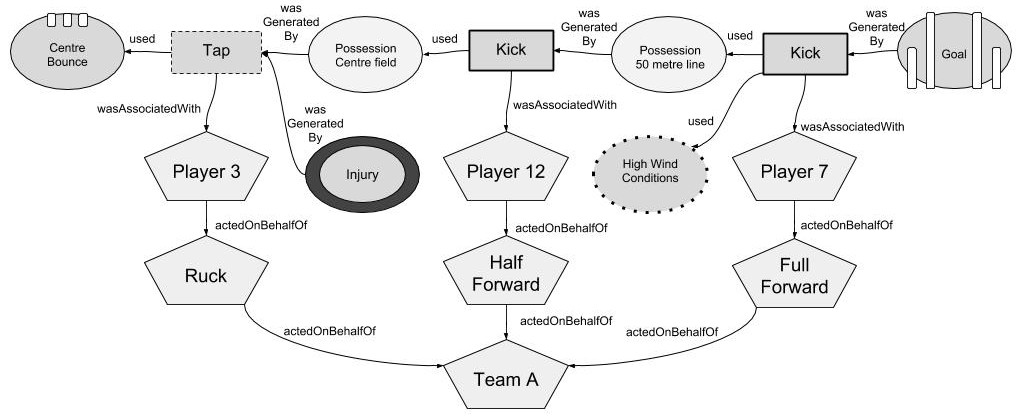}
\caption{Example use of our notation to describe the
physical provenance of a goal}
\label{PhysExample}
\end{figure}
}

{We provide an example of how the provenance of the goal described in
the Motivating Scenario could be modelled in Fig \ref{PhysExample}. We
see that the goal resulted from a kick performed by Player 7, who
possessed the ball as a result of a kick by Player 12, who in turn
possessed the ball as a result of a tap by Player 3 from the centre
bounce which served as the origin of the possession chain.}

{Due to the tendency of sport to focus on the single point of the ball,
we can see the provenance information tends to take the form of a
sequential chain. In a hypothetical variant of the game with multiple
balls, the provenance would take the form of a graph with parallel
branches for each ball and occasional cross-links when games events
relating to one ball interfere with game events relating to the other
ball. Nevertheless, our example still includes some branching, such as
injuries generated by game events that may be handled while the rest of
the game progresses, and external events such as wind conditions that
occasionally interact with the game through influencing the outcome of a
kick.}

\begin{figure}
\includegraphics[width=\columnwidth]{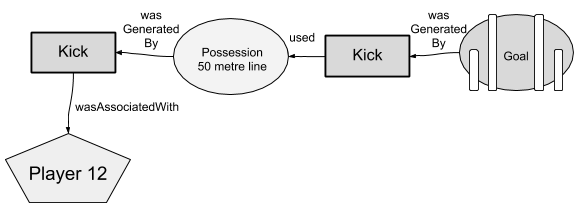}
\caption{Example provenance query answer returned to the user in our notation.}
\label{PhysQueryExample}
\end{figure}

{By annotating the game in such a manner, it becomes possible to express
queries about game events in the same manner as one would query a more
conventional data provenance graph. For example, the performance analyst
may be interested in how a goal came to be, specifically examining goal
assists. Without provenance, the performance analyst would have to
either rewatch the raw video for the game or read the match feed and
filter out irrelevant information. With provenance, they can query the
provenance graph for influences on the creation of the goal,
supplementing their query to filter to certain node types or depth
limits (in this case, filtering to chains involving agents separated by
2 activities). An example of the result one might receive is shown in
Fig \ref{PhysQueryExample}.}

{}

{}

\begin{table*}[!t]
\caption{Semantic constructs for provenance in the sports domain}
\label{Params}
\begin{tabular}{p{2cm}p{8cm}p{3cm}p{1cm}}
\toprule
\textbf{Semantic Construct (W3C PROV)} & \textbf{Description (in context of Sport)}                                                                             & \textbf{Specialised Semantic Construct  (in context of Sport)} & \textbf{ID} \\
\midrule
\multirow{3}{*}{Entity}      & \multirow{3}{8cm}{Entities can be either digital data, or physical concepts such as the state of having possession of the ball.} & Video feed                                                                                      & 1  \\
                             &                                                                                                                                & Physical game state                                                                             & 2  \\
                             &                                                                                                                                & Metric                                                                                          & 3  \\
\midrule
\multirow{3}{*}{Activity}    & \multirow{3}{8cm}{A process, whether manual or automated}                                                                        & Annotation                                                                                      & 4  \\
                             &                                                                                                                                & Computation                                                                                     & 5  \\
                             &                                                                                                                                & De-identify                                                                                     & 6  \\
\midrule
\multirow{5}{*}{Agent}       & \multirow{5}{8cm}{The person or device involved in performing an activity.}                                                      & Human                                                                                           & 7  \\
                             &                                                                                                                                & Player                                                                                          & 8  \\
                             &                                                                                                                                & Player Role                                                                                     & 9  \\
                             &                                                                                                                                & Sensor                                                                                          & 10 \\
                             &                                                                                                                                & Web portal                                                                                      & 11 \\
\midrule
\multirow{2}{*}{Connection}  & \multirow{2}{8cm}{While data provenance deals with data dependency, physical provenance deals with causality.}                   & Data dependency                                                                                 & 12 \\
                             &                                                                                                                                & Physical causality                                                                              & 13 \\
\bottomrule
\end{tabular}
\end{table*}

\subsection{Workflow Provenance}

{}

{In the previous section, we showed how the W3C PROV specification could
be translated into the sports domain to model physical provenance.
However, as most sport games focus on a single linear sequence of
events, representing the physical aspects of the game as a provenance
graph is, by itself, of limited benefit when compared to a traditional
linear timeline. The true value of this approach comes when provenance
can be traced throughout the entire system to link game events with
player metrics.}

{}

{In this section, we consider the use of the W3C PROV specification to
describe the derivation of digital data, such as metrics, computed as
part of a workflow. As this task is more abstract, the concepts at this
level are not clearly sports specific, especially when compared to our
physical provenance model for sport. Nevertheless, we argue that the
functional and quality requirements of the sports domain have
implications on the selection of an appropriate workflow
representation.}

{}

{Video analysis is one of the primary tools that sport performance
analysts use to analyse the game and communicate results to players and
coaches. This is evidenced by the popularity of video timeline based
annotation tools such as Sportscode amongst
elite sports teams. As such, our representation introduces a custom
symbol for video data, and we envision that if our representation was
used as part of an interactive tool, it would allow the user to directly
play back video segments when they form part of the provenance graph,
without the need to open the video in an external program and scan to
the time of events.}

{}

{Sport analysis workflows requires a combination of automated processing
(e.g. metric calculation) and manual processing (e.g. video annotation).
The W3C PROV standard does not make any distinction between manual
versus automated processes, so in theory can model both. However, in
practice, due to its generality, capturing automated processes fully
such that they could be recomputed requires extending the standard to
specify these details, such as to capture the source code and software
environment involved.}

\begin{figure}[!htb]
\includegraphics[width=0.8\columnwidth]{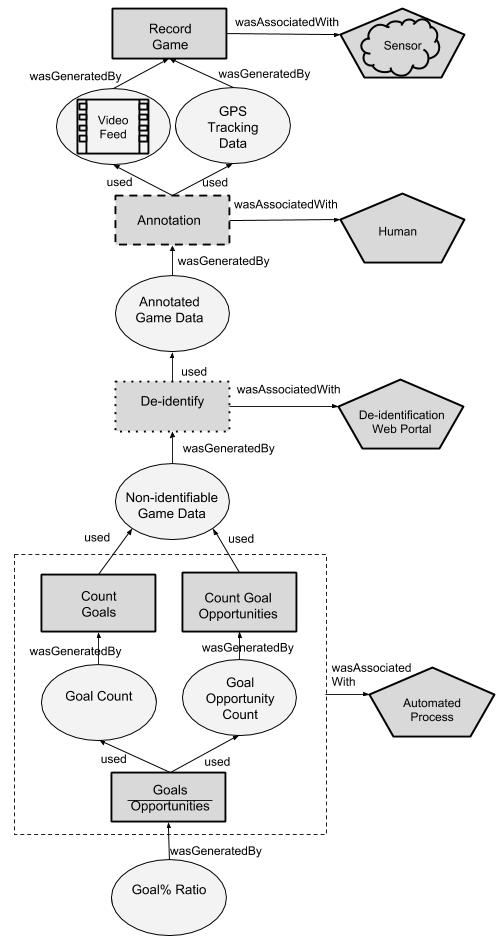}
\caption{Example use of our notation to capture to describe the data provenance of the Goal\% Ratio metric}
\label{WorkflowProv}
\end{figure}

Unlike the physical sciences, sports science involves working with
human participants (i.e. sports players). As such, there is often a need
to de-identify data for privacy reasons, for example, if a sports club
decides to share player data with researchers outside the club. This has
implications on the provenance capture system, as it means that
different users need access to different parts of the provenance graph
(e.g. the researcher should have an incomplete graph that prevents them
tracing provenance of the player data back past the de-identify
operation, while the sport club should be able to reconstruct the entire
provenance graph once the researcher shares their final findings and
provenance data).

In Fig \ref{WorkflowProv}, we present an example of our proposed
notation to capture the provenance of a computation of player goal
accuracy.

\subsection{Combined Provenance}

\begin{figure}[htbp]
\includegraphics[width=\columnwidth]{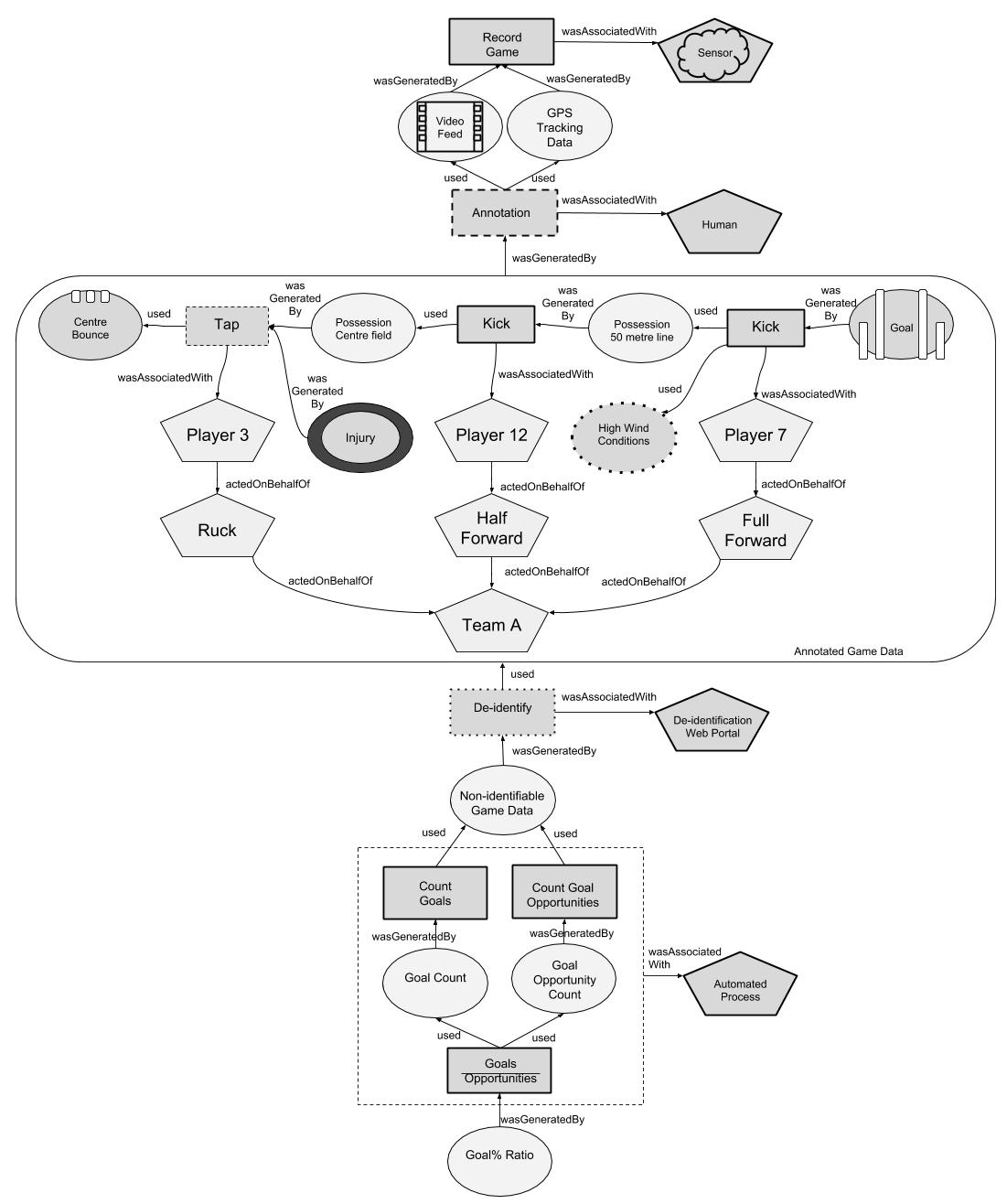}
\caption{Example use of our notation to describe physical and data provenance together as part of same provenance graph}
\label{CombinedProv}
\end{figure}

In the previous sections, we suggested notation for physical provenance
to describe game events and separately for workflow provenance to
describe metrics and computations. In Fig \ref{CombinedProv}, we show
that the annotated game dataset that forms part of the workflow can be
decomposed into the underlying game events it represents, and thus
physical provenance and workflow provenance can be integrated as part of
a single provenance graph.

Combining our customised notation for workflow and physical provenance
graphs ensures that all aspects of the provenance system will be
expressed using concepts the user can interpret. For example, consider
that a sport performance analyst performs a query to trace the
provenance of a metric back the the game events that contributed to it.
While the query references a metric (Goal\% Ratio, etc.) that is defined
at the workflow level, the resulting answer needs to be in terms of game
events, which can be communicated in the language of sport practitioners
by using the same physical provenance notation used to express the
physical query response in Fig \ref{PhysQueryExample}. This prevents the
user from being exposed to the underlying system encoding of the game
data (as would be the case if they exported the game events using an
arbitrary format determined by their video annotation software), thus
increasing the overall usability of the system through consistency and
familiarity of the representation.

While the broad semantic constructs such as Entities, Activites,
Agents, and Connections already exist in the W3C PROV standard, we
highlighted the need for specialised semantic constructs (along with
syntactic representations) to meet the needs of the sports domain. We
provide an overview of the key specialisations required in Table
\ref{Params}.


\section{Comparative Evaluation}

In this section we compare our proposed approach to the W3C PROV
standard and the VisTrails workflow management system, within the
context of the sports domain. We evaluate their functionality against
the tasks outlined in our motivating scenario (section \ref{sec:provmotivation}), the effectiveness of
their visualisation against design principles described by the Physics
of Notations framework \cite{Moody2010}, and their usability against Nielsen's heuristics
for user interface design \cite{Nielsen1994}.

\begin{figure}[b]
\includegraphics[width=0.8\columnwidth]{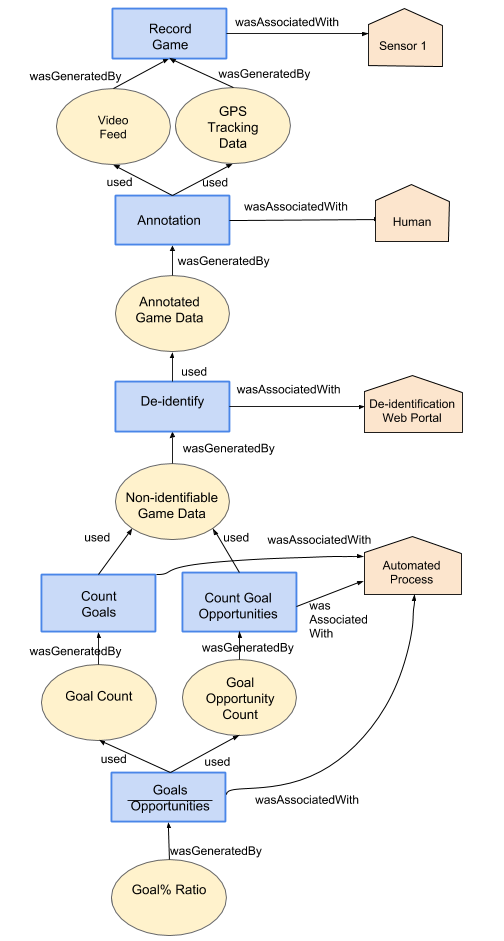}
\caption{Description of how Goal\% ratio was determined,
expressed using the W3C PROV standard. Note that the W3C PROV standard
only captures the activities and datasets at a high level and captures
neither the details of the dataset nor code necessary to reproduce the
process.}
\label{W3CPROV}
\end{figure}

\begin{figure}[b]
\includegraphics[width=0.9\columnwidth]{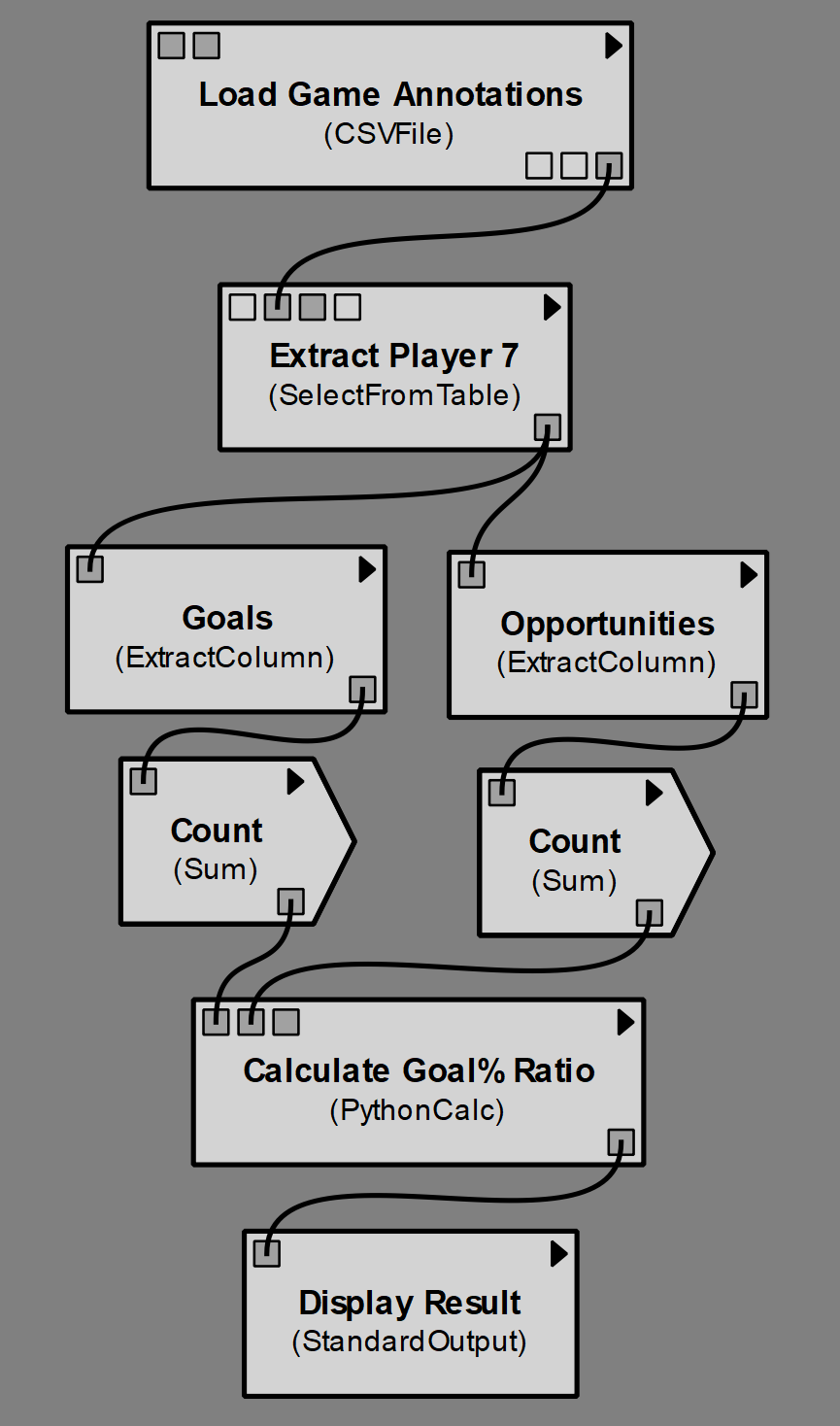}
\caption{Construction of pipeline to determining Goal\%
ratio of player using the Vistrails workflow system. Note that it is
not possible to describe the manual annotation processes in VisTrails,
so this has to be performed using an external system then loaded as the
first step of the pipeline.}
\label{VisTrails}
\end{figure}

\subsection{Functionality}

We will begin by modelling the workflow provenance scenario (see
Motivating Scenario) in each system so that we can compare differences
of the modelling languages.

The W3C PROV standard includes semantic constructs for modelling
entities (e.g. a dataset), activities (e.g. a process) and agents (e.g.
people that perform the process). It also includes the concept of a plan
to describe how a process was carried out, but the details of how to
execute a plan is left open, so cannot fully capture the details of an
automated process without introducing additional semantics. We use the
W3C PROV standard to describe the computation of player evaluation
metric in Fig \ref{W3CPROV}.

VisTrails models workflows as a directed graph of automated processing
elements (usually visually represented as rectangular boxes). Each
processing element has ``ports'' that represent the inputs (top of box)
and outputs (bottom of box) to/from the process. The user drags
connections between output ports and input ports to wire up the
workflow. Ports contain type information, which the interface uses to
prevent the user from accidentally connecting ports with conflicting
types. The resultant workflow is fully automated and reproducible,
however is not able to model processes that require human input, other
than at the level of tracking manual changes to the the workflow itself.
We show an implementation of the metric computation pipeline within
VisTrails in Fig \ref{VisTrails}.

We evaluate these systems against the requirements set out in the
Motivating Scenario.

\textbf{Integrated support for working with video data:} The W3C PROV standard
does not provide a means to directly represent datasets other than
as plain text using the \textit{prov:value} property. However, it integrates
with semantic web technologies such as the Resource Description
Framework (RDF) which could, in theory, be used to model and describe a
video source. VisTrails contains predefined modules for working with
tabular data, but does not provide inbuilt modules for working with
video data. One could implement custom modules for loading video data
and visualising the final output as video. However, without
architectural changes to the source code, the system does not have the flexibility
to support interactive editing or display of video sources as it flows
through the processing pipeline.

\textbf{Support for automated processes:} The W3C PROV standard includes the
concept of a plan to describe how an activity was conducted, but does
not capture details such as the source code or software environment that
would be needed to reproduce the process. In contrast, VisTrails is a
workflow automation tool designed to ensure reproducibility (although
this reproducibility may still be undermined by missing data or
dependencies on broken web services) and provides a selection of
built-in processing modules as well as allowing user-defined Python
scripts to cater to situations where the built-in modules are
insufficient for a particular task. VisTrails supports export to the W3C
PROV standard, but achieves this through mixing in resources within the
VisTrails namespace so that it can represent the concepts missing from
the PROV standard, as shown in the sample displayed in Listing
\ref{VTPROV}.

\begin{figure}[h] 
\begin{lstlisting}[frame=single, basicstyle=\small, caption={Sample of PROV export generated
by VisTrails. Note that it mixes resources in the Vistrails ``vt:'' namespace with the
W3C ``prov:'' namespace to make capturing the workflow possible.}, label=VTPROV]
<prov:document version="1.0.4"
  xmlns:dcterms="http://purl.org/dc/terms/"
  xmlns:prov="http://www.w3.org/ns/prov#"
  xmlns:vt="http://www.vistrails.org/registry.xsd">
  <prov:entity prov:id="e15">
    <prov:type>vt:data</prov:type>
    <prov:label>str_expr</prov:label>
    <prov:value>(player,==,7)</prov:value>
    <vt:id>15</vt:id>
    <vt:type>(org.vistrails.vistrails.basic:String,
              org.vistrails.vistrails.basic:String,
              org.vistrails.vistrails.basic:String)
    </vt:type>
    <vt:desc>(None,None,None)</vt:desc>
  </prov:entity>
  ...
</prov:document>
\end{lstlisting}
\end{figure}

\textbf{Support for manual interaction:}{~Because the W3C PROV standard does
not distinguish between manual and automated processes, and only
models details of activities at a high level, it is well suited to
describing manual processes and the agents (people) involved. VisTrails
provides a way for users to explore the parameter space and to
interactively view the output of the workflow, however does not provide
a way to capture manual processes as steps of the workflow, other than
by capturing the history of changes to the structure of the workflow
itself. Other workflow systems such as Taverna support interactive
processes as components of the workflow that either run locally and
interact with the user, or run through a web interface.\footnote{\url{https://taverna.incubator.apache.org/documentation/interaction/}} However these are limited to
self-contained sequential tasks rather than iterative ad-hoc tasks that
require interaction with the rest of the pipeline.}

{}

\textbf{Partial / shared workflow graphs:}{~The W3C PROV standard was designed
for sharing of provenance information on the web. References to
resources that make up the provenance graph are represented as }{URIs,
and thus information referenced by the provenance graph could
potentially be restricted by controlling access to the resources
referred to. As a concrete example, part of the provenance graph could
include a URI referencing a document that contains the mapping of player
identifiers to anonymised codes, however the document the URI refers to
could be hosted on the sport club's intranet and require a password to
gain access. Social platforms for scientific data sharing have proposed
sharing data alongside workflow information, such as MyExperiment
\cite{DeRoure2009} for sharing Taverna workflows, and CrowdLabs
\cite{Mates2011} for sharing VisTrails workflows. However, a study of
Taverna workflows shared on myExperiment found that ``nearly 80\% of the
tested workflows failed to be either executed or produce the same
results'' \cite{Zhao2012}, thus suggesting there still exist practical
issues sharing and archiving workflows in a manner that results can be
replicated, particularly in cases where certain data cannot be shared
for confidentiality reasons. VisTrails contains in-built support for
workflow ``diff'' and ``merge'', as well as ``visualisation by analogy''
which automatically translates changes applied to one workflow to
another workflow. These features could potentially ease collaboration on
shared workflows.}

{}

\textbf{Provenance / Reverse Debugging:}{~There are multiple types of
provenance information. ``Workflow provenance'' tracks the the processes
applied to datasets, but usually does not allow inspection of these
processes, whereas ``data provenance'' is fined-grained provenance that
tracks how individual data items are derived from each other
\cite{Tan2007}. Data provenance is further split into ``why'' provenance
\cite{Widom2000} which captures all data records that contribute to a
result, ``where'' provenance \cite{Buneman2001} which deals with only the
parts of records that are copied into a result, and ``dependency''
provenance \cite{Cheney2007}\cite{Cheney2011} which is similar to why
provenance, but formalises the notion of what it means for part of a
data record to contribute to a result.}

{}

{While VisTrails' provenance browser by default only shows
coarse-grained workflow provenance information pertaining to when each
component of the workflow was executed, the user can roll back to any
version of the workflow, modify the components of interest to output
additional debugging information such as inputs and outputs, then re-run
workflow using cached results where available. The W3C PROV standard
only deals with modelling and representing provenance, not how to
capture provenance. The level of granularity expressed is the choice of
the person or process that generates the provenance.}

{}

\textbf{Streaming data:}{~The W3C PROV standard can be used to describe
provenance in situations involving real-time streams of sensor data by
using the standard to describe the provenance of each individual sensor
observation \cite{Compton2014}. The VisTrails user manual includes a
section ``streaming in VisTrails'' that describes how functions can
incrementally process data. This could potentially be utilised to
process a stream of sensor data, however the stream would need to
terminate eventually for the workflow execution to complete
successfully.}

{}

{We summarise our above findings in Table \ref{EvalFunctionality}.}

\begin{table}[h]
  \caption{Evaluation of functionality against the
tasks outlined in Motivating Scenario}
  \label{EvalFunctionality}
\begin{minipage}{\columnwidth}
\begin{tabular}{ccc}
\toprule
\begin{minipage}[t]{0.30\columnwidth}\raggedright\strut
{\textbf{Requirement}}
\strut\end{minipage} &
\begin{minipage}[t]{0.30\columnwidth}\raggedright\strut
{\textbf{W3C PROV\footnotemark}}
\strut\end{minipage} &
\begin{minipage}[t]{0.30\columnwidth}\raggedright\strut
{\textbf{VisTrails}}
\strut\end{minipage}\tabularnewline
\midrule
\begin{minipage}[t]{0.30\columnwidth}\raggedright\strut
{Integrated support for working with video data}
\strut\end{minipage} &
\begin{minipage}[t]{0.30\columnwidth}\raggedright\strut
{No\footnotemark}
\strut\end{minipage} &
\begin{minipage}[t]{0.30\columnwidth}\raggedright\strut
{No\footnotemark[\value{footnote}]}
\strut\end{minipage}\tabularnewline
\midrule
\begin{minipage}[t]{0.30\columnwidth}\raggedright\strut
{Support for automated processes}
\strut\end{minipage} &
\begin{minipage}[t]{0.30\columnwidth}\raggedright\strut
{No\footnotemark[\value{footnote}]}
\strut\end{minipage} &
\begin{minipage}[t]{0.30\columnwidth}\raggedright\strut
{Yes}
\strut\end{minipage}\tabularnewline
\midrule
\begin{minipage}[t]{0.30\columnwidth}\raggedright\strut
{Support for manual interaction}
\strut\end{minipage} &
\begin{minipage}[t]{0.30\columnwidth}\raggedright\strut
{Yes}
\strut\end{minipage} &
\begin{minipage}[t]{0.30\columnwidth}\raggedright\strut
{Partial}
\strut\end{minipage}\tabularnewline
\midrule
\begin{minipage}[t]{0.30\columnwidth}\raggedright\strut
{Partial / shared workflow graphs}
\strut\end{minipage} &
\begin{minipage}[t]{0.30\columnwidth}\raggedright\strut
{Yes}
\strut\end{minipage} &
\begin{minipage}[t]{0.30\columnwidth}\raggedright\strut
{Partial}
\strut\end{minipage}\tabularnewline
\midrule
\begin{minipage}[t]{0.30\columnwidth}\raggedright\strut
{Provenance / Reverse Debugging}
\strut\end{minipage} &
\begin{minipage}[t]{0.30\columnwidth}\raggedright\strut
{Yes}
\strut\end{minipage} &
\begin{minipage}[t]{0.30\columnwidth}\raggedright\strut
{Partial}
\strut\end{minipage}\tabularnewline
\midrule
\begin{minipage}[t]{0.30\columnwidth}\raggedright\strut
{Streaming data}
\strut\end{minipage} &
\begin{minipage}[t]{0.30\columnwidth}\raggedright\strut
{Yes}
\strut\end{minipage} &
\begin{minipage}[t]{0.30\columnwidth}\raggedright\strut
{Partial}
\strut\end{minipage}\tabularnewline
\bottomrule
\end{tabular}
\end{minipage}
\end{table}
\addtocounter{footnote}{-1}
\footnotetext{For W3C PROV, we evaluate the ability to model provenance information, however an external system would be needed to actually
capture the provenance information and explore it.}
\addtocounter{footnote}{1}
\footnotetext{Partial support may be possible via extending the language with
additional modules / semantics.}

{}

\subsection{Effectiveness of visual notation}

{}

In Table \ref{EvalNotation}, we summarise our findings of the effectiveness of the visual notation used by each system. As the W3C PROV standard provides textual serialisations such as XML, but does not formally specify a visual notation, we evaluate the (non-normative) visualisations the standard uses to document examples.

{}

\subsection{Heuristic Usability Evaluation}

{}

In Table \ref{EvalUsability}, we summarise the usability issues
identified in VisTrails as a result of a heuristic evaluation. We did
not attempt to evaluate the usability of the W3C PROV standard, as it
does not specify any particular implementation to create provenance
documents.

{}

\begin{samepage}
\section{Key Findings}

\begin{enumerate}
\item Automated workflow tools often lack support for capturing ad-hoc
manual processes that cannot be automated. Conversely, provenance
standards such as W3C PROV recognise the need to document the inputs and
procedures involved in ad-hoc manual processes, but lack semantics for
describing the code and execution environment necessary to reproduce
automated parts of the analysis. Supporting the needs of the sports
domain -- and other fields where manual and automated analysis are
intertwined -- requires combining these as part of a unified standard to
ensure a complete and reproducible capture of the analysis.

\item As automated workflow tools treat processes as black boxes with
limited traceability, their provenance logs typically only show basic
execution information such as the time the process ran and status of the
result. However, analysts in the sport domain require fine-grained data
provenance to trace results back to raw events. Although the black box
nature of workflows prevents support of ``why'' provenance and ``where''
provenance methods designed for analysing provenance of SQL query
results, we noted that workflows implicitly support a form of
retrospective investigation through the ability to roll back history and
recompute key processes with additional logging information or with
modified data inputs to observe the effects on the output. In cases
where capturing fine-grained provenance is not possible, we suggest that
workflow systems could support the user to retrospectively reason
about the likely provenance of data by guiding the user through the
procedure of retrospectively collecting intermediate states and
manipulating inputs to infer which data values had an impact on the
result of the process. This approach could also be used to support user
reasoning about provenance in workflows that involve complex
probabilistic processes (such as neural networks) by supporting the user
with the tools to rewind the process and ``prod'' at intermediate data
to understand what is most relevant (i.e. sensitivity analysis) and
whether expected properties hold (i.e. metamorphic testing) rather than
overloading the user with information about the computations carried
out.

\item Our analysis of the notations used shows poor utilisation of the
available design space. Notably the ``graphic economy'' of the systems
studied could be improved by utilising additional visual variables such
as texture to further distinguish symbols. As certain domains demand a
different set of semantic constructs to others (e.g. the reliance on
video annotation within the sport domain), we advocate for optimising
the visual notation for the domain. Translating abstract provenance
concepts into concrete concepts in the language of the domain would
reduce the number of usability issues faced by practitioners.
\end{enumerate}
\end{samepage}

\section{Conclusions}

While general purpose workflow managers and provenance notations exist,
we have demonstrated that these systems need extensions and
specialisations respectively in order to express the sport domain. Our
proposed notation demonstrates what such a language could look like in
the sport domain, however would need to be supplemented with tooling to
make this a reality.

Future work is needed to evaluate how potential users respond to our
proposed notation. A study by Bachour et al. in which a computer game
presented gamers with a visualisation inspired by the W3C PROV standard
suggests that non-expert users may be confused by the direction of the
arrows, as they are intuitively interpreted as data flow rather than
data dependency \cite{Bachour2015}. An empirical evaluation is needed to
detect whether similar issues also exist in the sport domain.

We speculate that usability issues arising from the use of general
provenance systems in the context of a domain with specialised needs and
terminology could be hindering the uptake of provenance systems despite
the widely recognised need for reproducible research. While we have
explored issues from the perspective of the sport domain, it is possible
that other scientific subfields could also benefit through the
introduction of customised provenance languages for their scientific
domain. Thus another avenue for future work is to use our methodology to
generate a family of provenance systems, each optimised for a particular
scientific domain.

%% file: appendix.tex
\appendix
\clearpage
\onecolumn
\section{Appendices}

{
\renewcommand{\arraystretch}{1.8} 
\begin{longtable}[c]{@{}lll@{}}
\caption{Effectiveness of visual notation against principles of Physics of Notations \cite{Moody2010}}
\label{EvalNotation}\tabularnewline
\hline
\endhead
\toprule
\begin{minipage}[t]{0.30\columnwidth}\raggedright\strut
{\textbf{Criterion}}
\strut\end{minipage} &
\begin{minipage}[t]{0.30\columnwidth}\raggedright\strut
{\textbf{W3C PROV}}
\strut\end{minipage} &
\begin{minipage}[t]{0.30\columnwidth}\raggedright\strut
{\textbf{VisTrails}}
\strut\end{minipage}\tabularnewline
\midrule
\begin{minipage}[t]{0.30\columnwidth}\raggedright\strut
{Semiotic Clarity}{\\
}{(fraction of semantic constructs in Table \ref{Params} mapped to
unique symbols)}
\strut\end{minipage} &
\begin{minipage}[t]{0.30\columnwidth}\raggedright\strut
{4/13}

{}

{Contains high level semantics for entity, activity, agent and
connection.}
\strut\end{minipage} &
\begin{minipage}[t]{0.30\columnwidth}\raggedright\strut
{3/13}

{}

{Metric (port), computation, data dependency (connection).}

{}

{No concept of agents. No ability to directly model real world. No
concept of }{connection }{causality.}
\strut\end{minipage}\tabularnewline
\begin{minipage}[t]{0.30\columnwidth}\raggedright\strut
{Perceptual Discriminabilit}{y}{~\\
}{(fraction of symbols }{with unique visual variables}{)}
\strut\end{minipage} &
\begin{minipage}[t]{0.30\columnwidth}\raggedright\strut
{4/4}

{}

{Could be improved: different }{colours / shapes}{~for specialisations.
(Points still awarded because top level constructs have distinct
symbols)}
\strut\end{minipage} &
\begin{minipage}[t]{0.30\columnwidth}\raggedright\strut
{3/3}

{}

{Ports and activities share same shape as each other, but differ by
size.}

{}

{Could be improved: ports with different types should have different
colours / shapes. Activities with different types should have different
colours and use a larger variety of shapes. (Points still awarded for
these because only one type of sport semantic construct was supported)}
\strut\end{minipage}\tabularnewline
\begin{minipage}[t]{0.30\columnwidth}\raggedright\strut
{Semantic Transparency\\
}{(fraction of symbols with obvious meanings)}
\strut\end{minipage} &
\begin{minipage}[t]{0.30\columnwidth}\raggedright\strut
{0/4}

{}

{Use of circles for entities and rectangles for processes conflicts with
data flow diagrams (which use circles for processes). Use of house
shaped pentagons for agents is only memorable when agent represents an
organisation. Arrows are in direction of data dependency, but intuitive
interpretation is in direction of data flow.}
\strut\end{minipage} &
\begin{minipage}[t]{0.30\columnwidth}\raggedright\strut
{3/3}

{}

{Analogy: electric circuit (rectangular components, small contacts,
connection wires)}

{}

{Could be improved: While obvious square is a port, not obvious which
port is which (user has to memorise order). While obvious that box is a
process, specific type of process is not obvious (e.g. uses pentagon for
control flow rather than conventional diamond for ``if'' condition)}
\strut\end{minipage}\tabularnewline
\begin{minipage}[t]{0.30\columnwidth}\raggedright\strut
{Complexity Management\\
}{(can it visualize complex workflows?)}
\strut\end{minipage} &
\begin{minipage}[t]{0.30\columnwidth}\raggedright\strut
{Yes}

{}

{Ontologies support the ``Open-world assumption'', thus allowing
specifying as much or as little detail as appropriate.}
\strut\end{minipage} &
\begin{minipage}[t]{0.30\columnwidth}\raggedright\strut
{Yes}

{}

{Supports grouping nodes}
\strut\end{minipage}\tabularnewline
\begin{minipage}[t]{0.30\columnwidth}\raggedright\strut
{Cognitive Integration}{\\
}{(can the user navigate without getting lost?)}
\strut\end{minipage} &
\begin{minipage}[t]{0.30\columnwidth}\raggedright\strut
{Yes}

{}

{Includes concept of ``bundles'' to annotate information required to
navigate documents at meta-level. E.g. to describe provenance of
provenance information.}
\strut\end{minipage} &
\begin{minipage}[t]{0.30\columnwidth}\raggedright\strut
{Yes}

{}

{Top level workflow acts as overview, then user can drill down into
parameter values, history variations, etc.}
\strut\end{minipage}\tabularnewline
\begin{minipage}[t]{0.30\columnwidth}\raggedright\strut
{Visual Expressiveness}{\\
(fraction of visual variables used)}
\strut\end{minipage} &
\begin{minipage}[t]{0.30\columnwidth}\raggedright\strut
{2/8}

{}

{Shape and colour.}
\strut\end{minipage} &
\begin{minipage}[t]{0.30\columnwidth}\raggedright\strut
{1/8}

{}

{Shape}

{}

{Colour is used for execution state, but this is not one of semantic
constructs, and brightness is used to determine if a port is connected,
but neither of these map to semantic constructs of relevance.}
\strut\end{minipage}\tabularnewline
\begin{minipage}[t]{0.30\columnwidth}\raggedright\strut
{Dual Coding}{\\
(fraction of symbol parameters with multiple unique visual variables)}
\strut\end{minipage} &
\begin{minipage}[t]{0.30\columnwidth}\raggedright\strut
{1/3}

{}

{Shape and colour used together to ensure symbols are distinct (i.e.
colours improve distinguishability of symbols, and even if user is
colour blind, symbols are still distinguishable by shape)}
\strut\end{minipage} &
\begin{minipage}[t]{0.30\columnwidth}\raggedright\strut
{0/3}

{}

{In theory shape and colour can be assigned if designing custom module,
but colour is not used in any of the default modules.}
\strut\end{minipage}\tabularnewline
\begin{minipage}[t]{0.30\columnwidth}\raggedright\strut
{Graphic Economy}{\\
(total symbols, less is better as it reduces cognitive load)}
\strut\end{minipage} &
\begin{minipage}[t]{0.30\columnwidth}\raggedright\strut
{4}
\strut\end{minipage} &
\begin{minipage}[t]{0.30\columnwidth}\raggedright\strut
{3}

{}

{(If we were to remove all features that we are not assessing)}
\strut\end{minipage}\tabularnewline
\begin{minipage}[t]{0.30\columnwidth}\raggedright\strut
{Cognitive Fit}{\\
(is the notation understandable to performance analysts?)}
\strut\end{minipage} &
\begin{minipage}[t]{0.30\columnwidth}\raggedright\strut
{Partial}

{}

{When arrows are labeled, visual notation is unambiguous.}
\strut\end{minipage} &
\begin{minipage}[t]{0.30\columnwidth}\raggedright\strut
{Partial}

{}

{Intuitive flow metaphor, however advanced operations require writing
custom Python scripts.}
\strut\end{minipage}\tabularnewline
\bottomrule
\end{longtable}
}

\clearpage
{
\renewcommand{\arraystretch}{1.8} 
\begin{longtable}[c]{@{}lll@{}}
\caption{Usability evaluation of VisTrails using
Nielsen's top ten heuristics \cite{Nielsen1994}}
\label{EvalUsability}\tabularnewline
\hline
\endhead
\toprule
\begin{minipage}[t]{0.30\columnwidth}\raggedright\strut
{\textbf{Criterion}}
\strut\end{minipage} &
\begin{minipage}[t]{0.30\columnwidth}\raggedright\strut
{\textbf{Support}}
\strut\end{minipage} &
\begin{minipage}[t]{0.30\columnwidth}\raggedright\strut
{\textbf{Issues}}
\strut\end{minipage}\tabularnewline
\midrule
\begin{minipage}[t]{0.30\columnwidth}\raggedright\strut
{Visibility of system status}
\strut\end{minipage} &
\begin{minipage}[t]{0.30\columnwidth}\raggedright\strut
{Shows progress indicator when evaluating workflow. Displays which
modules executed / have errors.}
\strut\end{minipage} &
\begin{minipage}[t]{0.30\columnwidth}\raggedright\strut
{}
\strut\end{minipage}\tabularnewline
\begin{minipage}[t]{0.30\columnwidth}\raggedright\strut
{Match between system and the real world}
\strut\end{minipage} &
\begin{minipage}[t]{0.30\columnwidth}\raggedright\strut
{Boxes for processes connected by lines resembles real-world electronic
wiring of modules.}
\strut\end{minipage} &
\begin{minipage}[t]{0.30\columnwidth}\raggedright\strut
{Some terms may present confusion for non-technical users:
``PythonCalc'' (evaluate an expression), ``StandardOutput'' (display
result in the terminal), and ``Map'' (a higher order function, not a
geological map).}
\strut\end{minipage}\tabularnewline
\begin{minipage}[t]{0.30\columnwidth}\raggedright\strut
{User control and freedom}
\strut\end{minipage} &
\begin{minipage}[t]{0.30\columnwidth}\raggedright\strut
{Full tracking of history as tree}
\strut\end{minipage} &
\begin{minipage}[t]{0.30\columnwidth}\raggedright\strut
{}
\strut\end{minipage}\tabularnewline
\begin{minipage}[t]{0.30\columnwidth}\raggedright\strut
{Consistency and standards}
\strut\end{minipage} &
\begin{minipage}[t]{0.30\columnwidth}\raggedright\strut
{}
\strut\end{minipage} &
\begin{minipage}[t]{0.30\columnwidth}\raggedright\strut
{Some terms such as ``port'' (rather than input / output) may increase
time to learn.}
\strut\end{minipage}\tabularnewline
\begin{minipage}[t]{0.30\columnwidth}\raggedright\strut
{Error prevention}
\strut\end{minipage} &
\begin{minipage}[t]{0.30\columnwidth}\raggedright\strut
{Ports have types to ensure that user can only connect two ports if
their types match.}
\strut\end{minipage} &
\begin{minipage}[t]{0.30\columnwidth}\raggedright\strut
{}
\strut\end{minipage}\tabularnewline
\begin{minipage}[t]{0.30\columnwidth}\raggedright\strut
{Recognition rather than recall}
\strut\end{minipage} &
\begin{minipage}[t]{0.30\columnwidth}\raggedright\strut
{The system provides some support to aid the user's memory (e.g. dark
ports to remind the user a default has been set)}
\strut\end{minipage} &
\begin{minipage}[t]{0.30\columnwidth}\raggedright\strut
{The user needs to memorise}{~}{the port order of modules to use the
interface efficiently.}
\strut\end{minipage}\tabularnewline
\begin{minipage}[t]{0.30\columnwidth}\raggedright\strut
{Flexibility and efficiency of use}
\strut\end{minipage} &
\begin{minipage}[t]{0.30\columnwidth}\raggedright\strut
{Provides shortcut key combinations for advanced users}
\strut\end{minipage} &
\begin{minipage}[t]{0.30\columnwidth}\raggedright\strut
{}
\strut\end{minipage}\tabularnewline
\begin{minipage}[t]{0.30\columnwidth}\raggedright\strut
{Aesthetic and minimalist design}
\strut\end{minipage} &
\begin{minipage}[t]{0.30\columnwidth}\raggedright\strut
{Main focus of the application is on the workflow}
\strut\end{minipage} &
\begin{minipage}[t]{0.30\columnwidth}\raggedright\strut
{}
\strut\end{minipage}\tabularnewline
\begin{minipage}[t]{0.30\columnwidth}\raggedright\strut
{Help users recognize, diagnose, and recover from errors}
\strut\end{minipage} &
\begin{minipage}[t]{0.30\columnwidth}\raggedright\strut
{System highlights module(s) with error}
\strut\end{minipage} &
\begin{minipage}[t]{0.30\columnwidth}\raggedright\strut
{Use of colour as sole indicator of error could be problematic for users
with colour blindness.}
\strut\end{minipage}\tabularnewline
\begin{minipage}[t]{0.30\columnwidth}\raggedright\strut
{Help and documentation}
\strut\end{minipage} &
\begin{minipage}[t]{0.30\columnwidth}\raggedright\strut
{User manual includes step-by-step guidelines on how to use.}

{}

{In-built option to display documentation for the selected module}
\strut\end{minipage} &
\begin{minipage}[t]{0.30\columnwidth}\raggedright\strut
{In-built documentation for module often missing}
\strut\end{minipage}\tabularnewline
\bottomrule
\end{longtable}
}